\documentclass[twocolumn,showpacs,preprintnumbers,amsmath,amssymb,
superscriptaddress, groupedaddress]{revtex4}

\usepackage{graphicx}
\usepackage{dcolumn}
\usepackage{bm}

\begin{document}

\title{Absence of Wavepacket Diffusion in Disordered Nonlinear Systems}

\author{G. Kopidakis}
\affiliation{Max Planck Institute for the Physics of Complex Systems, 
N\"othnitzer Str. 38, D-01187 Dresden, Germany}
\affiliation{Department of Materials Science and Technology,
University of Crete, 71003 Heraklion, Greece}
\author{S. Komineas}
\affiliation{Max Planck Institute for the Physics of Complex Systems, 
N\"othnitzer Str. 38, D-01187 Dresden, Germany}
\author{S. Flach}
\affiliation{Max Planck Institute for the Physics of Complex Systems, 
N\"othnitzer Str. 38, D-01187 Dresden, Germany}
\author{S. Aubry}
\affiliation{Max Planck Institute for the Physics of Complex Systems, 
N\"othnitzer Str. 38, D-01187 Dresden, Germany}
\affiliation{Laboratoire L\'eon Brillouin (CEA-CNRS), CEA Saclay,
91191-Gif-sur-Yvette, France}

\date{\today}

\begin{abstract}
We study the spreading of an initially localized  wavepacket  in two nonlinear chains 
(discrete nonlinear Schr\"odinger and quartic Klein-Gordon) with disorder. 
Previous studies suggest    
that there are many initial conditions such that the second moment of the norm and
energy density distributions diverge as a function of time. 
We find that the participation number of a wavepacket does not diverge simultaneously.
We prove this result analytically for norm-conserving models
and strong enough nonlinearity.
After long times the dynamical state consists of a distribution of nondecaying yet interacting normal modes.
The Fourier spectrum shows quasiperiodic dynamics.
Assuming this result holds for any initially localized wavepacket, a limit profile for the 
norm/energy distribution with infinite second moment should exist in all cases which rules out the 
possibility of slow energy diffusion (subdiffusion). This 
limit profile could be a quasiperiodic solution (KAM torus).
\end{abstract}

\pacs{Valid PACS appear here}
\maketitle

It is well-known that Anderson localization occurs for a one-dimensional linear system
with uncorrelated random potential. 
Since  all the linear eigenmodes -- Anderson modes (AMs) -- are localized,  
any wavepacket which is initially localized
remains localized for all time. Therefore there is no energy diffusion \cite{PWA58}.
When nonlinearities are added to such models, AMs interact with each other, giving rise
to more complex situations \cite{GrKiv92}. 
Numerical studies of wavepacket propagation in several models 
showed that the second moment of the norm/energy 
distribution growths subdiffusively in time as $t^{\alpha}$ \cite{Shep93, Mol98,PS07},
with $\alpha$ in the range $0.3-0.4$, though not being accurately 
determined. 
The conclusion was that the initial excitation will completely delocalize for infinite times. 
Recently, experiments were performed on light
propagation in spatially random nonlinear optical media \cite{Exp}.

Spatially periodic nonlinear systems will support Discrete Breathers (DBs),
which are spatially localized time periodic solutions \cite{ST88}
with frequencies outside the frequency spectrum of the linear system. 
The temporal evolution of a localized wavepacket 
leads to the formation of a DB, while a part of the energy of the wavepacket
is radiated ballistically to infinity (in the form of weakly nonlinear plane waves) \cite{DBWP}. 
In that case, the second moment of the energy density distribution diverges 
as $t^2$, falsly suggesting complete delocalization.
The  participation number $P$ of the norm/energy distribution (or similar quantities)
is a well-known measure of the degree of localization.
In the case of a periodic nonlinear lattice, $P$ will saturate at a finite value,
correctly indicating the formation of a DB.

For nonlinear random systems it was proven rigorously that  
AMs survive in the presence of nonlinearities as spatially localized and 
time-periodic solutions \cite{AF91} with frequencies which depend on the amplitude of the mode. 
The allowed frequencies form a fat Cantor set (with finite measure) whose
density becomes unity for weak nonlinearity.
They are located inside the frequency spectrum of the linear system.
Numerical techniques for obtaining these (dynamically stable) intraband DB solutions at computer 
accuracy were developed \cite{KA99}.
When they are chosen as an initial wavepacket,
they persist for infinite time and there is no diffusion at all. 

Here we analyse carefully the evolution of the participation number of wavepackets
as a function of time,
in situations where subdiffusion is claimed to exist \cite{Shep93, Mol98,PS07}. 
We study two models.  
The Hamiltonian of the disordered discrete nonlinear Schr\"odinger equation (DNLS) 
\begin{equation}
\mathcal{H}_{D}= \sum_n \left(\epsilon_n |\psi_n|^2-\frac{1}{2} \beta |\psi_n|^4-V ( \psi_{n+1}\psi_{n}^{\star}
+\psi_{n+1}^{\star}\psi_{n} )\right)
\label{RDNLS}
\end{equation}
with complex variables $\psi_n$. The random on-site energies $\epsilon_n$ are  
chosen uniformly from the interval $\left[-\frac{W}{2},\frac{W}{2}\right]$.  
The equations of motion are generated by $\dot{\psi}_n = \partial \mathcal{H}_{D}/
\partial (i \psi^{\star}_n)$.
We choose $\beta=1$ and $V=-1$ here \cite{Mol98} and
note that varying the norm of the initial wavepacket 
is strictly equivalent to varying $\beta$.

The Hamiltonian of the quartic Klein-Gordon chain (KG) 
\begin{equation}
\mathcal{H}_{K}= \sum_n  \frac{p_n^2}{2} +\frac{\tilde{\epsilon}_n}{2} u_n^2 + 
\frac{1}{4} g u_n^4+\frac{V}{2}(u_{n+1}-u_n)^2\;.
\label{RQKG}
\end{equation}
The equations of motion are $\ddot{u}_n = - \partial \mathcal{H}_{K} /\partial u_n$,
$\tilde{\epsilon}_n= 1+\epsilon_n(W=1)$, and $g=1$. 

For $\beta=g=0$ both models are reduced to the linear eigenvalue problem
$\lambda A_n = \epsilon_n A_n -V(A_{n+1} + A_{n-1})$. The eigenvectors $A^{\nu}_n$
are the AMs, and the eigenvalues $\lambda_{\nu}$ are the frequencies of the AMs
for the DNLS, while the KG modes have frequencies $\omega_{\nu}=\sqrt{\lambda_{\nu} +1+2V}$.
   
Hamiltonian (\ref{RDNLS}) (unlike (\ref{RQKG})), in addition to conserving the energy,
also conserves the total norm $S=\sum_n |\psi_n|^2=\langle\psi|\psi\rangle$. We use this norm conservation 
for proving  rigorously that initially localized wavepackets 
with a large enough amplitude  cannot spread  to arbitrarily small amplitudes. 
The consequence is that  a part of the initial energy must remain well-focused at all times. 

This proof is inspired by \cite{RABT00}.  We split the total energy 
$\mathcal{H}_{D}= \langle\psi|\mathbf{L}|\psi\rangle + 
H_{NL}$ 
into the sum of its  
quadratic term of order $2$ and its nonlinear terms of order strictly higher than $2$.
Then, $\mathbf{L}$ is a linear operator which is bounded from below (and above).
In our specific example, we have  $\langle\psi|\mathbf{L}|\psi\rangle \geq \omega_{m} 
\langle\psi|\psi\rangle= \omega_{m} S$ 
where $\omega_{m} \geq  -2 -\frac{W}{2}$ is the lowest eigenvalue of $\mathbf{L}$.  
Otherwise, the higher order nonlinear terms have to be strictly negative.

If we assume that the wavepacket amplitudes spread to zero at infinite time,  we have
$\lim_{t\rightarrow +\infty}( \sup_n  |\psi_n|)=0$. 
Then 
$\lim_{t\rightarrow +\infty} ( \sum_n  |\psi_n|^4) <  \lim_{t\rightarrow +\infty}
( \sup_n |\psi_n^2|) (\sum_n  |\psi_n|^2)  =0$ since  $S=\sum_n \psi_n^2$ is time invariant.  
Consequently ,  for $t \rightarrow +\infty$ we have $\mathcal{H}_{NL}=0$ and $\mathcal{H}_D 
\geq  \omega_{m} \sum_n |\psi_n|^2=\omega_{m}  S$. 
Since $\mathcal{H}_D$ and $S$ are both time invariant, this inequality should be fulfilled at all times.
However when  the initial amplitude $A$ of the wavepacket is large enough, it cannot be 
initially fulfilled  because the nonlinear energy diverges as $-A^4$  while 
the total norm  diverges as  $A^2$ only.  For example, a wavepacket initially at a 
site $0$ ( $\psi_n=0$ for $n\neq 0$ and $\psi_0 =\sqrt{A}$ ) has energy $\mathcal{H}_D=  
\epsilon_0 A^2  - \frac{1}{2} A^4$.
Consequently, the above inequality is not fulfilled when  $A^2> -2 (\omega_m - \epsilon_0)>0$.
Thus such an initial wavepacket cannot spread to zero amplitudes at infinite time.

This proof is valid for DNLS models with any $W$ (including the periodic case)
and any lattice dimension and can be easily 
extended to larger classes of DNLS models where the nonlinear terms are either strictly 
negative, or strictly positive.  
Note that the large amplitude regime where we prove that 
complete energy diffusion is impossible in DNLS models, 
is precisely the one where subdiffusion is claimed to completely delocalize the wavepacket  
\cite{PS07}. Thus we disprove these claims.  

We performed extensive numerical simulations, and characterized the wavepacket spreading
both in real space for DNLS and normal mode space (Anderson space or AS) for KG. 
We used initial wavepackets with all the
energy localized on a single site $n_0$, or single AM, or combinations, close to $n_0$.
Nonlinearity induces diffusion in Anderson space,
where each AM is characterized by a amplitude $a_{\nu}$ and momentum $\dot{a}_{\nu}$.
We analyze distributions $z_l \geq 0$ using the second moment
$m_2= \sum_l (l-l_0)^2 z_l$ and the participation number $P=(\sum_l z_l )^2 / \sum_l z_l^2$,
which measures the number of the strongest excited sites in $z_l$.
We order the AMs in space by increasing value of the center-of-norm coordinate $X_{\nu}=\sum_n n A_n^2$.
In the results presented here, for the DNLS $z_n=|\psi_n|^2$ is the norm density in real space, 
and for the KG $z_{\nu}=\dot{a}_{\nu}^2/2+\omega_{\nu}^2a_{\nu}^2/2$ is the (harmonic)
energy density in AS. The system size was $N=1000$ for KG, and $N=2000$ for DNLS.
Excitations did not reach the boundaries during the integration time, and results are 
unchanged when further increasing $N$.

\begin{figure}
\includegraphics[angle=-90,width=1.25\columnwidth]{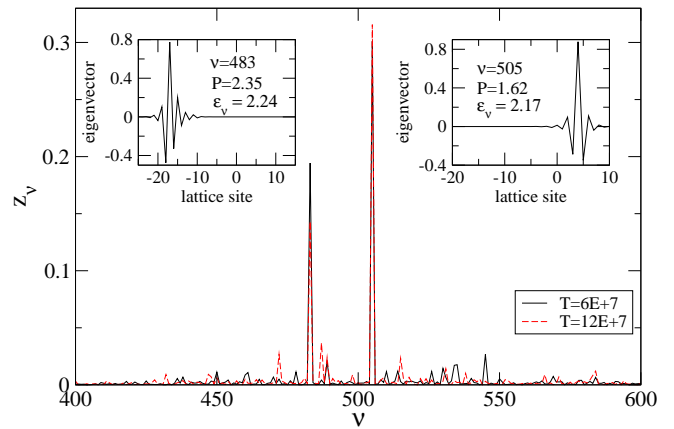}
\caption{(color online) KG: Energy distribution at $t=6\times 10^7$
(black solid) and $t=1.2 \times 10^8$ (red dashed) in AS. Initial single site excitation
with energy $E=1$, $V=0.25$. 
Insets: profiles of the strongest excited AMs in real space.}
\label{figsk1}
\end{figure}

\begin{figure}
\includegraphics[angle=-90,width=1.25\columnwidth]{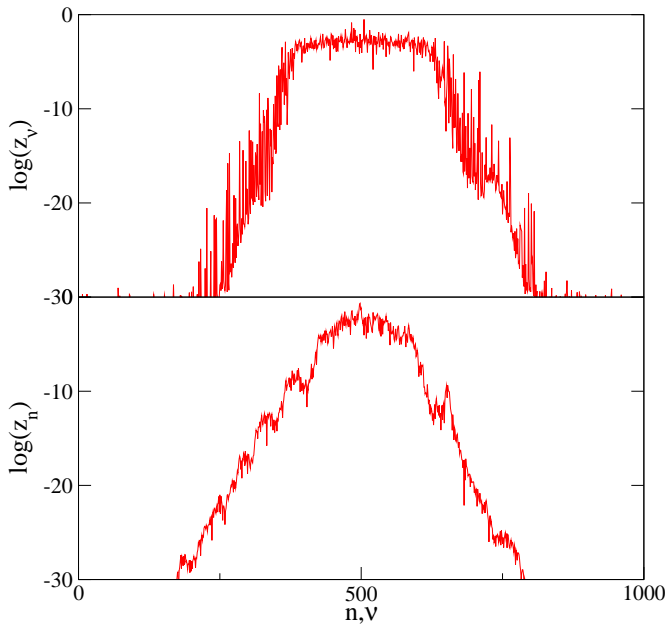}
\caption{(color online) Same as in Fig.\ref{figsk1} but on a logarithmic scale.
Top panel: KG, $t=1.2 \times 10^8$, AS.
Bottom panel: DNLS, $W=4$,  $t=1.2 \times 10^8$, real space.}
\label{figsk2}
\end{figure}

\begin{figure}
\includegraphics[angle=-90,width=1.25\columnwidth]{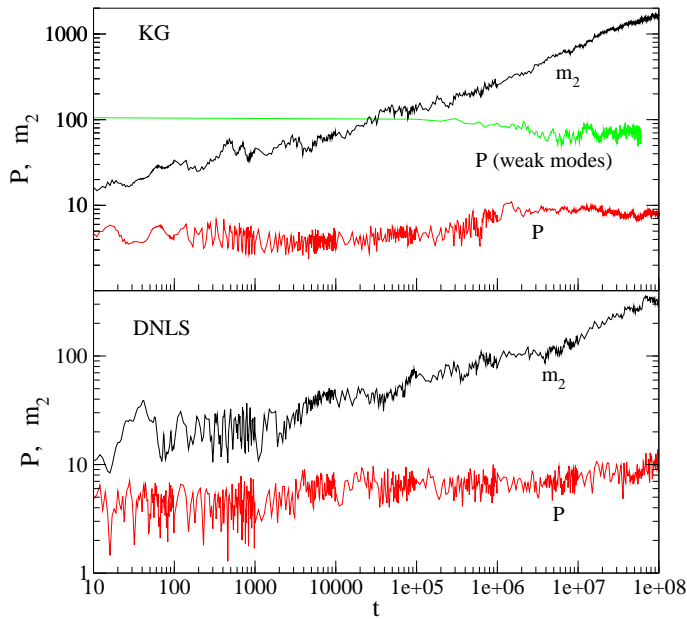}
\caption{(color online) $P$ and $m_2$ versus time, on logarithmic scale. 
Parameters as in Fig.\ref{figsk2}. Top panel: KG, AS. Bottom panel: DNLS, real space.}
\label{figsk3}
\end{figure}

We show in Fig.\ref{figsk1} the KG energy distribution in AS for a single
site excitation with energy $E=1$, and $V=0.25$ at times $t=6\times 10^7,1.2 \times 10^8$. 
Two rather strongly excited
modes are surviving almost unchanged on these time scales.
The insets show their eigenvectors, which are well localized, and practically
do not overlap. 
The same distributions on a logarithmic scale (KG and DNLS) show a chapeau of weaker excited AMs,
with exponential tails due to its finite width (Fig.\ref{figsk2}). 
This chapeau is perhaps slowly
growing. The subdiffusive growth of the second
moment at these times (see Fig.\ref{figsk3}) is mainly due to weak excitation of tail modes.

The participation number $P(t)$ is plotted in Fig.\ref{figsk3} for the same runs. 
We observe {\sl no} growth. $P$ fluctuates around a value of 7-10,
confirming the results in Fig.\ref{figsk1}, that we observe a localized state, similar to
a DB. 
Assume that the rest of the weakly excited modes continues to subdiffuse in the chapeau.
We use a modified distribution $z_{\nu}$ for the KG run, where the 10 strongest mode
contributions are zeroed (top panel, green curve). 
The weak mode participation number
is now fluctuating around 70, but {\sl again does not grow}. Therefore the chapeau 
appears not to diffuse, and the observed growth of $m_2\sim t^{0.3...0.4}$ 
is not related to a delocalization process. Instead, we find that the packet {\sl does not
delocalize}. Indeed, assuming that the chapeau homogeneously spreads in a subdiffusive way
as claimed, it follows that $P(t) \sim t^{\alpha/2}$, which clearly contradicts our observations.
We repeated these runs with various initial conditions and disorder realizations 
with similar results. However the localization pattern (Fig.\ref{figsk1}),
and the observed averaged participation number $P$, fluctuate. Performing an averaging
of the final distribution over several realizations \cite{Shep93, Mol98,PS07} 
will therefore completely smear out
the sharp localization patterns in the distributions. 
Closer inspection of the evolution of $m_2$ shows, that the exponent $\alpha$ 
is strongly depending on the time intervals of study, 
and also on the given disorder realization. There are  some indications 
suggesting that  $\alpha$ might decay at long time  and even that $m_2(t)$ may  saturate,
but further clarification may call for very extensive numerical investigations.

\begin{figure}
\includegraphics[angle=-90,width=1.25\columnwidth]{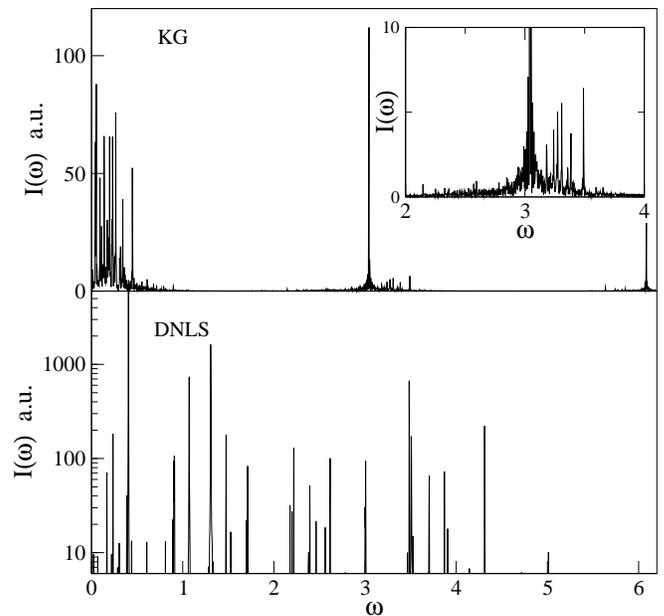}
\caption{(color online) $I(\omega)$ for KG and DNLS.
Parameters as in Fig.\ref{figsk3}. Insert: magnification of the main peak.}
\label{figsk4}
\end{figure}

Finally we calculated the Fourier transform $I(\omega)$ of $P(t)$ (after $t=2\times 10^7$, over an interval of
$\Delta t = 2000$), see Fig.\ref{figsk4}. We find a quasiperiodic spectrum, which
is close to periodic, with no hints of a chaos-induced continuous part.
For the KG case the energy densities are quadratic forms of the AM coordinates, thus the
main peak position $\omega \approx 3$ corresponds to a frequency $\omega \approx 1.5$ for the
AM coordinate dependence, which coincides with the frequencies of the strong excited Anderson modes
in Fig.\ref{figsk1}.

Our main result is, that in both models (\ref{RDNLS}) and (\ref{RQKG}), whatever the 
initial wavepacket is (even if it is not fulfilling the conditions for our theorem), 
and irrespective of the model parameters and the 
disorder realization, the participation number  does not diverge as a function of time  
as it should  in case of subdiffusion (as $t^{\alpha/2}$)  but instead 
fluctuates between finite upper and lower bounds. 

Let us now propose an interpretation of our observations. First, it is useful to
recall the wavepacket behavior in the absence of disorder. When its amplitude is large enough for generating a   DB, there is a transient  dynamical state which is more or less chaotic, with a broad band time-Fourier 
spectrum overlapping  the spectrum of the linear system. 
Because of that, a part of the energy of the 
wavepacket is radiated to infinity. 
With that, the remaining DB like excitation becomes quasiperiodic first, and finally,
approaches an equidistant spectrum of periodic motion, which completely stops further
radiation.
The energy which has been emitted spreads towards infinity.   
Therefore there is a limit profile which is a localized time-periodic solution -  
an exact DB. This is the only possibility for the limit profile, in order to avoid
radiation. This is an example where the initial wavepacket self-organizes in order to stop 
radiation. 

When the system is both random and nonlinear, radiation into the linear spectrum is impossible
due to Anderson localization. Nevertheless, the same process starts as before, but  the energy emitted   by the initial wavepacket cannot spread towards infinity since the participation number (full and partial) does not diverge. The following cascading scenario may be true.
The core of the wavepacket emits a part of its energy which remains within 
the linear localization length nearby the initial wavepacket  
(due to the nonlinearity-induced coupling between the AMs).  
The same process should 
repeat for the emitted energy. A part of it remains localized while  another part 
is  reemitted  a bit farther from the central site within the localization length and so on.  This process 
of reemission repeats forever and generates a  tail for the wavepacket which will become 
much  more extended than the localization length. The central amplitude of the wavepacket does not 
tend to zero. The process of energy reemission  slows down when  the amplitude at the edge of the tail 
becomes small which explains the very slow numerical convergence. 
 The final result is that at infinite time, the energy (or norm) distribution should converge to a 
 nonvanishing limit profile which is summable since energy (or norm) is conserved. However,
  it may or may  not have a  finite second moment, 
  which makes the question of the evolution of the second moment secondary.
  Unlike the standard DB case in spatially periodic systems, the limit profile is not a
  time periodic solution.
 
It was proven rigorously (\cite{FSW86,XY02}) that stable spatially 
localized quasiperiodic solutions with finite energy exist in similar nonlinear models with infinitely many 
degrees of freedom without or with degenerate linear spectrum.  These KAM tori 
are quasiperiodic DBs which  in some sense are linear combination of Anderson modes 
surviving in the presence of nonlinearity.  
Indeed, we find that the Fourier spectrum of the wavepacket dynamics
becomes quasiperiodic, with narrow peaks and a small background as time grows 
suggesting  the motion tends to become \textit{quasiperiodic} (Fig.\ref{figsk4}).  

If the limit profile becomes a KAM torus, we should also observe that the largest Lyapunov 
exponent tends to $0$ as  $t\rightarrow +\infty$.  
Indeed, we find that this Lyapunov exponent drops rapidly during the first expansion part of
the wavepacket, and slowly further decays, with characteristic values of $10^{-4}$ at the end
of our simulations. The corresponding time scale is $10^4$, and four orders of magnitude smaller
than the simulation times. No chaotic dynamics is observable, and we
think that the convergence to the final KAM torus is very slow
 because the surrounding KAM tori are expected to become dense. We should even expect to
 enter the regime of Arnol'd diffusion which is 
 expected to be very slow and difficult to investigate both numerically and 
 analytically.
 
 Note that this convergence to a quasiperiodic limit profile can only 
 occur in infinite systems because if the system is finite, the regularization process of the 
 initially chaotic trajectories ends when the packet tails reach the edge of the box. Then, 
 we should expect to get equipartition of the energy after a sufficiently 
 large time and a trajectory which 
 remains chaotic with a nonzero largest Lyapunov exponent.

In summary, we have proved  by a rigorous analytical argument, and completed by numerical 
investigations of the participation number,  that a wavepacket in a random nonlinear system 
does not spread ad infinitum. A limiting quasiperiodic profile is approached, and the 
slow increase of the second moment of the energy/norm distribution does not violate
these findings. It is an open question whether the limiting profile will have a finite or
infinite second moment.
Thus, we observe absence of diffusion in nonlinear disordered systems.
Note that this conclusion can be equally well applied to higher dimensional systems,
provided all AMs are localized.

\begin{acknowledgments}
This work was performed within the program of the
Advanced Study Group 2007 at the MPIPKS Dresden
\texttt{http://www.mpipks-dresden.mpg.de/~asg2007/}.
 G.K and S.A  acknowledge the MPIPKS for its hospitality
 and also Egide and the Greek G.S.R.T. for support through Platon program 
(2007-2008).
 We thank V. Fleurov, R. Schilling, L. S. Schulman,
 and D. Shepelyansky  
 for useful discussions.
\end{acknowledgments}


\end{document}